\documentclass[conference]{IEEEtran}
\IEEEoverridecommandlockouts

\usepackage{cite}
\usepackage{amsmath,amssymb,amsfonts}
\usepackage{algorithmic}
\usepackage{graphicx}
\usepackage{textcomp}
\usepackage{xcolor}
\def\BibTeX{{\rm B\kern-.05em{\sc i\kern-.025em b}\kern-.08em
    T\kern-.1667em\lower.7ex\hbox{E}\kern-.125emX}}

\usepackage{subcaption}
\usepackage{multirow}

\begin{document}

\title{Targeted Data Poisoning for Black-Box Audio Datasets Ownership Verification}

\author{\IEEEauthorblockN{Wassim (Wes) Bouaziz}
\IEEEauthorblockA{\textit{Meta, FAIR} \\
\textit{CMAP, École polytechnique} \\
Paris, France \\
wesbz@meta.com}
\and
\IEEEauthorblockN{El-Mahdi El-Mhamdi}
\IEEEauthorblockA{\textit{CMAP, École polytechnique}\\
Palaiseau, France}
\and
\IEEEauthorblockN{Nicolas Usunier}
\IEEEauthorblockA{\textit{Meta, FAIR} \\
Paris, France}
}

\maketitle

\begin{abstract}
Protecting the use of audio datasets is a major concern for data owners, particularly with the recent rise of audio deep learning models.
While watermarks can be used to protect the data itself, they do not allow to identify a deep learning model trained on a protected dataset.
In this paper, we adapt to audio data the recently introduced \textit{data taggants} approach.
Data taggants is a method to verify if a neural network was trained on a protected image dataset with top-$k$ predictions access to the model only.
This method relies on a targeted data poisoning scheme by discreetly altering a small fraction (1\%) of the dataset as to induce a harmless behavior on out-of-distribution data called \textit{keys}.
We evaluate our method on the Speechcommands and the ESC50 datasets and state of the art transformer models, and show that we can detect the use of the dataset with high confidence without loss of performance.
We also show the robustness of our method against common data augmentation techniques, making it a practical method to protect audio datasets.
\end{abstract}


\section{Introduction}
Availability of audio datasets is crucial for the development of deep learning models.
The lack of transparency from model developers regarding their training data is however a major concern for data owners.
Protecting audio datasets against unauthorized use is thus important to ensure the integrity and ownership of the data.
While watermarks can be used to protect the data itself, they do not allow for dataset ownerhsip verification (DOV) which aims at identifying if a deep learning model was trained on a protected dataset.
Existing methods for DOV rely mostly on backdoor data poisoning \cite{Wenger2022DataIF,Li2022UntargetedBW,li2023black}, by having a model change its behavior when its inputs $x$ are perturbed with a specific pattern $x+t$.
These methods not only conflict with adversarial robustness, but they also cannot guarantee that a benign model will not be affected by the trigger pattern, thus providing no guarantee on the false positive rate.\\
In this paper, we build on \textit{Data taggants} \cite{datataggants}, a method to verify if a black-box neural network was trained on a protected dataset and only requires access to top-$k$ predictions of the model.
This method relies on a targeted data poisoning scheme 
to induce a harmless behavior on out-of-distribution secret data called \textit{keys}.
We adapt this notion of \textit{keys} to audio data modality.
We evaluate our method on the Speechcommands and the ESC50 datasets and state of the art transformer models and show that we can detect the use of the protected dataset with high confidence even at 1\% contamination rate while conserving the model's performances.
We also show the robustness of our method against common data augmentation techniques, making it a practical method to protect audio datasets.

\begin{figure}[t]
    \centering
    \includegraphics[width=0.483\textwidth]{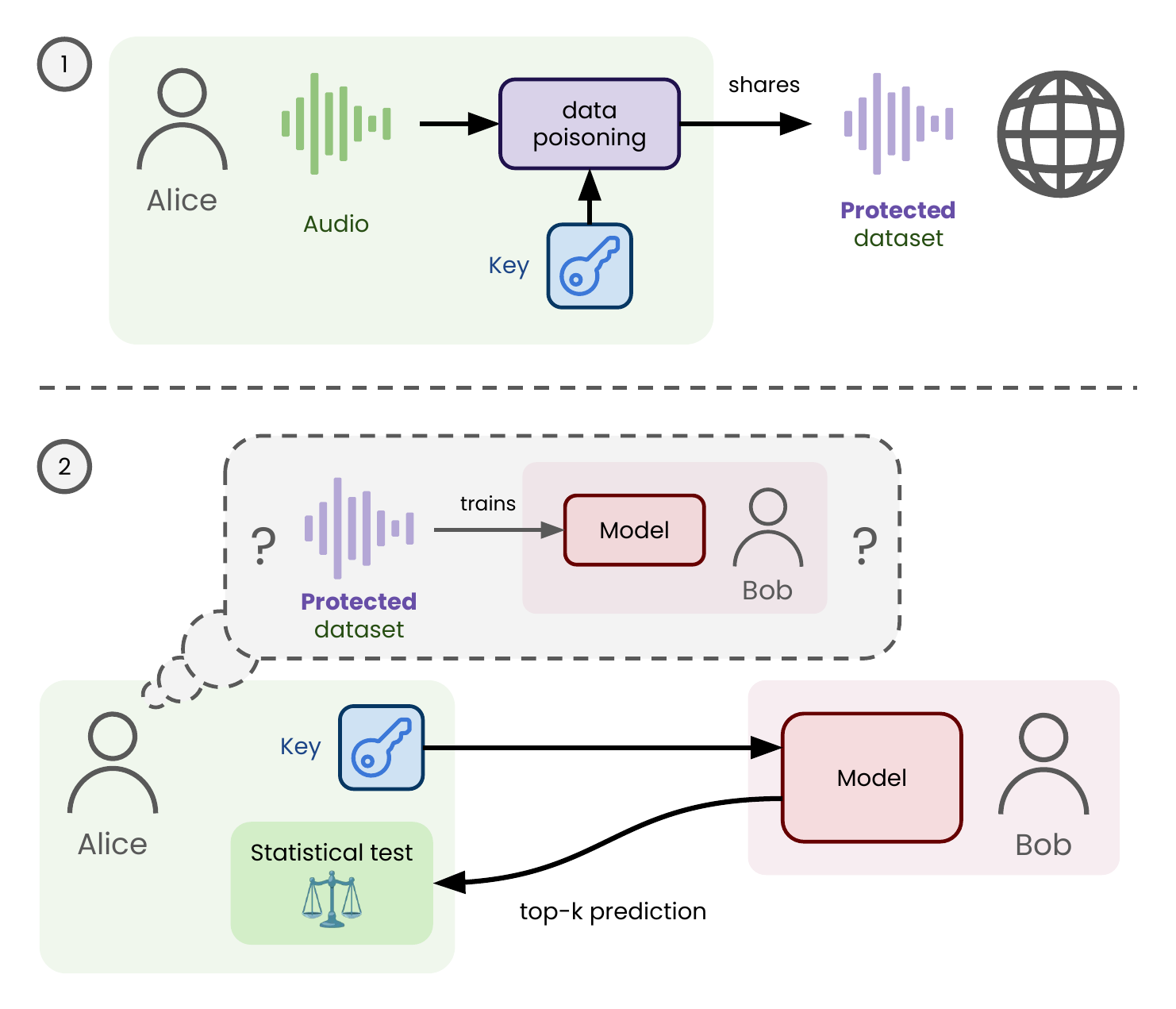}
    \caption{Overview. 1. Alice, the dataset owner, poisons a fraction of her dataset to induce a particular behavior on Bob's model.
    2. She can verify if Bob's model was trained on her dataset by using the top-$k$ predictions on a set of keys.}
    \label{fig:overview}
\end{figure}

\paragraph*{Setting} Alice, a dataset provider, wants to share her audio dataset but suspects that Bob, a model developer, will try to use her dataset to train a deep learning model without her approval.
We consider a practical black-box setting where Alice only has access to Bob's model's top-$k$ predictions.
Most importantly, Alice needs a method that is: \textit{stealthy} to avoid being detected by Bob; \textit{integrity-preserving} to allow authorized actors to legitimately use the dataset; and \textit{robust} to common data augmentation techniques and transformations that Bob is likely to apply during training.

\section{Related work}

\subsection{Data poisoning}
A critical threat to the integrity of machine learning models is data poisoning, where an attacker injects malicious data in the training set to manipulate the model's behavior at test time.
We distinguish between \textit{targeted data poisoning} where the attacker aims to manipulate the model's behavior on a specific set of inputs, and \textit{backdoor data poisoning} where the attacker aims to manipulate the model's behavior on inputs perturbed with a specific trigger pattern \cite{vassilev2024adversarial}.
Gradient alignment is a successful approach in both targeted and backdoor data poisoning \cite{geiping2020witches,Souri2021SleeperAS}.
It relies on crafting imperceptible perturbations in the input space such that information about the desired behavior is hidden in the gradients computed on the perturbed inputs.

\subsection{Audio watermarking}
Watermarking is the process of marking data with a message related to that data, such as a proof of ownership~\cite{cox2007digital}.
Its use on audio data for copyright protection has been extensively studied, with traditional signal processing techniques: least significant bit \cite{cvejic2004increasing}, echo hiding \cite{gruhl1996echo,cho2004echo}, patchwork \cite{arnold2000audio,yeo2003modified,kang2008full}, spread spectrum \cite{bender1996techniques,cox1997secure,Neoliya2002DIGITALAW} and quantization index modulation \cite{chen2001quantization}.
More recently, deep learning approaches have been proposed~\cite{singh2024silentcipher}, in particular for AI generated audio \cite{chen2023wavmark,roman2024proactive}.
Although these methods are efficient to protect the data itself, including against adversarial attacks \cite{chen2023wavmark}, and even if these methods radiate through generative models when the whole dataset is watermarked \cite{roman2024latentwatermarkingaudiogenerative}, they do not come with a systematic method to identify a deep learning model trained on a protected dataset.

\subsection{Dataset ownership verification}
The problem of verifying if a machine learning model has been trained on a protected dataset has mainly been studied in the context of image data.
Most approaches rely on backdoor attacks \cite{Wenger2022DataIF,Li2022UntargetedBW,li2023black}, where a small fraction of the dataset is poisoned such that a trained model behaves significantly differently on a benign input $x$ 
 and $x+p$, with a specific pattern $p$.
Backdoor-based approaches thus directly oppose to adversarial robustness, a desirable property for deep learning models, and were shown to be harmful to the then trained models \cite{Guo2023DomainWE}.
\cite{kim2020digital} proposed a method analogous to a visual watermark for audio data: they apply a filter to the magnitude of the short time Fourier transform of the audio signal.
A non-backdoor-based method proposed by \cite{Guo2023DomainWE} relies on crafting hard examples for models to predict and generate a data poisoning targeting these hard examples to facilitate their memorization.
Another non-backdoor approach on image datasets \cite{datataggants} targets particular data samples called \textit{keys}, crafted to have guarantees against false positives as opposed to previous methods without degrading the models' performances.

\section{Targeted data poisoning for dataset ownership verification}

\subsection{Motivation}
We adapt the data taggants \cite{datataggants} approach, introduced for image datasets, to audio datasets.
Data taggants rely on crafting out-of-distribution samples called \textit{keys} such that a benign model's behavior can be characterized on the keys.
We craft the \textit{keys} to be out-of-distribution data points.
For instance, Figure~\ref{fig:key_generation} shows, for a sound classification task, a spectral representation of such \textit{keys}, which ought to be points that do not correspond to any actual audio sample.

\subsection{Data taggants}
We consider Alice's dataset $D_{\mathcal{A}} = \{x_i, y_i\}_{i=1}^N$ where $x_i$ is an audio sample and $y_i$ its corresponding label.
Alice will poison a fraction $\epsilon$ of her dataset with perturbations $\Delta = \{\delta_i\}_{i=1}^N$ to obtain a poisoned dataset $\hat{D}_{\mathcal{A}} = \{x_i + \delta_{i}, y_i\}_{i=1}^N$ on which Bob will train his model $f(\cdot, \theta)$ parametrized by $\theta$.
Data taggants \cite{datataggants} rely on inducing a particular behavior from Bob's model on a set of $K$ keys $D_{K} = \{x^{(key)}_i, y^{(key)}_i\}_{i=1}^K$: each key $x^{(key)}_i$ must obtain a specific prediction $y^{(key)}_i$.
Alice can only tamper with her data \textit{once} before sharing them, and \textit{cannot} know Bob's parameters.
The data poisoning must thus be robust to different initializations.
Alice needs to craft the perturbations $\Delta$ such that learning on her dataset also forces Bob to learn the keys:
\begin{align*}
    \min_{\Delta} & \sum_{i=1}^{K} L(f(x^{(key)}_i, \theta(\Delta)), y^{(key)}_i) \text{ s.t.} \\
    \theta(\Delta) & \in \arg\min_{\theta} \sum_{i=1}^{N} L(f(x_i + \delta_i, \theta), y_i)
\end{align*}
$L$ is the loss function used to train Bob's model.
Similarly to \cite{geiping2020witches,datataggants}, we use \textit{gradient alignment} to achieve this goal.
Alice uses a surrogate model $f(\cdot, \theta_{\mathcal{A}})$ trained on her clean dataset, and crafts the perturbations of the poisoning set taken from the power set over the training data indices $S_{P} \in P([1..N])$ s.t. $\frac{|S_{P}|}{N} = \epsilon$ as to minimize the alignment loss:
\begin{align*}
    \min_{\Delta} \frac{1}{K} \sum_{i=1}^{K} \cos \bigg( & \nabla_{\theta} L (f(x^{(key)}_{i}, \theta_{\mathcal{A}}), y^{(key)}_{i}), \\
    & \sum_{j \in S_{P_{i}}} \nabla_{\theta} L (f(x_{j} + \delta_{j}, \theta_{\mathcal{A}}), y_{j}) \bigg)
\end{align*}
With $S_{P} = \cup_{i} S_{P_{i}} = \cup_{i} \{j; j \in S_{P} \wedge y_{j} = y^{(key)}_{i} \}$.
We solve this optimization problem using signed Adam, clipping after each step to ensure low amplitude perturbations, like in~\cite{datataggants}, but removed the image perceptual loss term.

\begin{figure*}[h]
    \centering
    \includegraphics[width=0.88\textwidth]{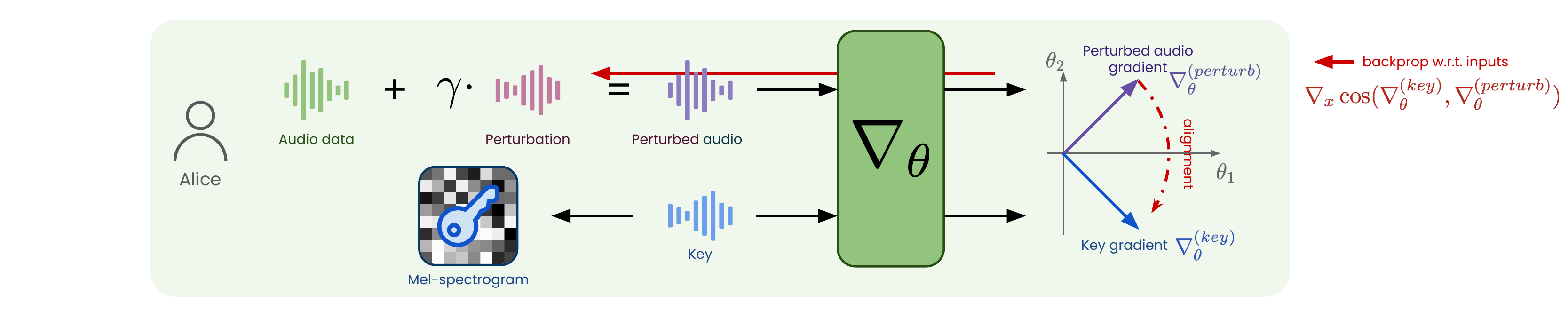}
    \caption{Illustration of the data poisoning process. Alice poisons a fraction $\epsilon$ of her dataset to align the gradients of perturbed data with the key gradient and induce a particular behavior on Bob's model.}
    \label{fig:poisoning}
\end{figure*}

\subsection{Detection}
To verify if Bob's model was trained on Alice's dataset, Alice uses the predictions of Bob's model over the set of keys.
A model trained on Alice's poisoned dataset $\hat{D}_{\mathcal{A}}$ should predict label $y^{(key)}_i$ for key $x^{(key)}_i$.
We use the same approach to detection and hypothesis test as \cite{datataggants} and consider the detection to be successful when a model displays a top-$k$ accuracy on the set of keys above a threshold $\tau > 0$.
We also need to ensure that a benign model, not trained on Alice's dataset, will not display the same behavior.
The choice for $k$ and $\tau$ balances between the false positive rate and the true positive rate (or detection rate).
The TPR is the probability of Bob's model to be correctly identified as being trained on $\hat{D}_{\mathcal{A}}$, and the FPR of its probability to be mistakenly identified as having been trained on $\hat{D}_{\mathcal{A}}$.

\paragraph*{Hypothesis testing} Under the null hypothesis $\mathcal{H}_{0}$: ``Bob's model was not trained on Alice's dataset'', given that the key labels $y^{(key)}_i \in \mathcal{Y}$ are attributed randomly, a benign model should perform at chance level on the set of keys.
$T_{k}$, the measured top-$k$ accuracy of such model should follow a binomial distribution with parameters $K$ and $k/C$ with $C = |\mathcal{Y}|$.
We can run a binomial test and compute its p-value with $p = \mathbb{P}(Z \geq T_{k}) = \sum_{z \geq T_{k}} \binom{K}{z} \left(\frac{k}{C}\right)^{z}\left(\frac{C-k}{C}\right)^{K-z}$, the probability of a random variable $Z$ following a binomial distribution with parameters $K$ and $k/C$ to be at least $T_{k}$.
This test only requires access to the top-$k$ predictions of Bob's model on the set of keys, which is a practical assumption.

\section{Experiments}

\subsection{Settings}
\subsubsection{Experimental details}
To demonstrate the effectiveness of our approach, we choose to experiment our approach on AST \cite{gong2021ast}, a transformer-based model trained on Speechcommands \cite{warden2018speech} or ESC50 \cite{piczak2015dataset} audio classification datasets.
We use a state of the art training recipe \cite{gong2021ast} using Adam optimizer, frequency and time masking, and mixup \cite{zhang2017mixup} data augmentations.
We consider a fraction $\epsilon = 1\%$ of the dataset to be poisoned and clip the perturbations waveform to a maximum amplitude of $0.05$, using order of magnitude less poisoned samples than \cite{kim2020digital}.
In each experiment, we sample $K = 10$ keys and use top-$k$ accuracy on this set with $k \in [1,10]$ to detect if Bob used Alice's dataset and derive a p-value.
We only consider here the case of Bob's model having the same architecture as Alice's model but with a different initialization, and trained on the poisoned dataset.
We repeat each experiment 5 times and combine the p-values using Fisher's method \cite{fisher1970statistical}.
We report the validation accuracy, the top-$k$ accuracy on keys, and the obtained p-values.
To preserve the utility of the data, the validation accuracy should not be significantly affected by the data poisoning.

\subsubsection{Choice of keys}
\begin{figure}
    \centering
    \includegraphics[width=0.25\textwidth]{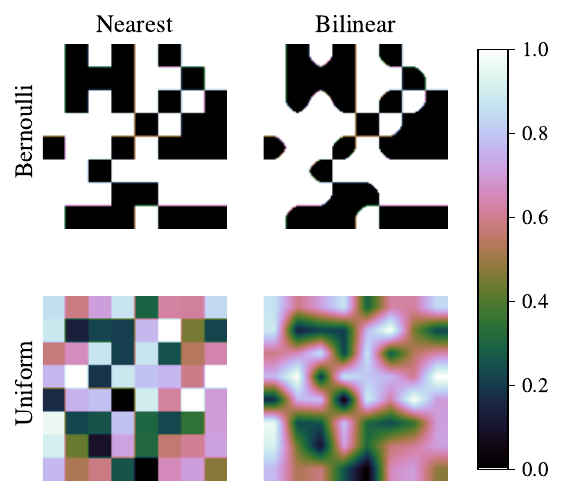}
    \caption{Illustration of the matrices $M$ obtained with the considered strategies of sampling and interpolation.}
    \label{fig:key_generation}
    \vspace{-0.2cm}
\end{figure}

Choosing out-of-distribution data as keys allows to limit the risk of interfering with the learner's objective and improve effectiveness of the detection \cite{datataggants}.
Since most audio models rely on spectral features, we generate the keys in the spectral domain.
We generate a $d \times d$ matrix $M$, resize it to fit with the model's input dimensions, and treat it as a mel-spectrogram of the $i$-th key $x^{(key)}_{i}$.
We apply the Griffin-Lim algorithm \cite{griffinlim} to reconstruct the audio signal from $M$ and use the obtained audio as the key $x^{(key)}_{i}$.
We generate $M$ by sampling either from a Bernoulli distribution or a $[0,1]$ uniform distribution, and we resize it either with nearest neighbor or bilinear interpolation.
Figure \ref{fig:key_generation} illustrates the considered samplings and interpolations for $d = 8$.

\subsection{Results}

\begin{figure}[h]   
    \begin{subfigure}{0.483\textwidth}
        \includegraphics[width=\textwidth]{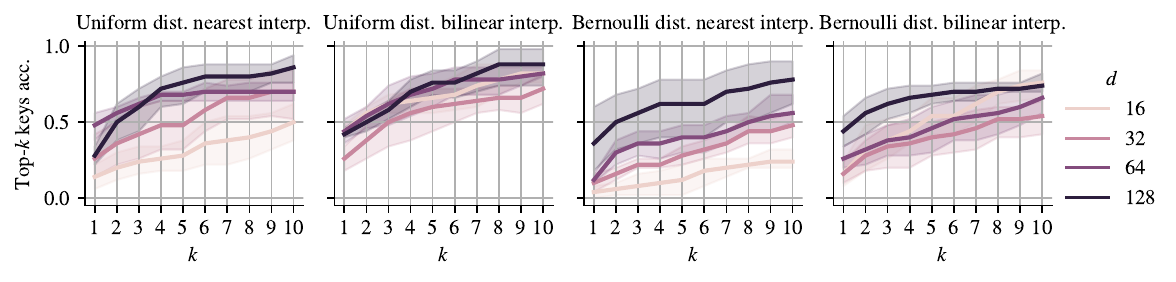}
        \caption{Top-$k$ accuracies on keys.}
        \label{fig:keys_acc}
    \end{subfigure}
    \begin{subfigure}{0.483\textwidth}
        \includegraphics[width=\textwidth]{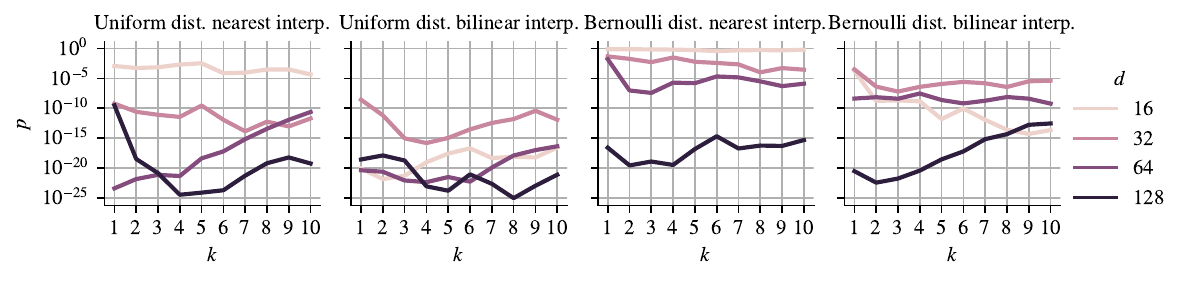}
        \caption{Associated $p$-values guarantee a low false positive rate even at low values of $k$.}
        \label{fig:p_values}
    \end{subfigure}
    \begin{subfigure}{0.483\textwidth}
        \includegraphics[width=\textwidth]{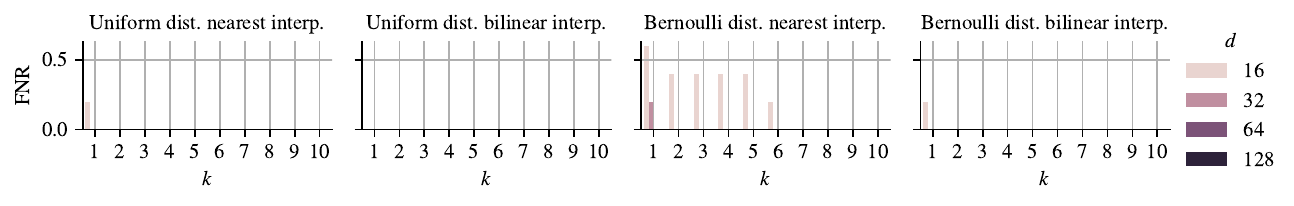}
        \caption{False Negative Rate shows that our method allows to consistently detect Bob's model except for $d=16$.}
        \label{fig:fnr}
    \end{subfigure}
    \caption{Results of the data poisoning on the Speechcommands dataset with the AST model.}
    \label{fig:results_speechcommands_ast}
\end{figure}

\paragraph*{Effectiveness \& Robustness} We evaluate the \textit{effectiveness} of our method with the top-$k$ accuracy on keys and associated p-value.
Using data augmentations and different initializations also allows us to evaluate the robustness of our approach.
Figure \ref{fig:results_speechcommands_ast} shows the top-$k$ accuracies on keys, corresponding $p$-values, and the measured false negative rate (FNR -- the ratio of models trained on the protected dataset to evade detection) for the Speechcommands dataset with the AST model and different keys generation strategies.
Notice that the top-$k$ accuracy on keys is not important per se.
What matters most is the associated $p$-value, which characterizes the FPR, the probability for a benign model to display the detected behavior.
Figure \ref{fig:p_values} shows that our method gives extremely low $p$-values for most keys generation strategies, \textbf{as low as} $\mathbf{10^{-25}}$ for uniform distribution with bilinear interpolation and $d = 128$.
Figure \ref{fig:fnr} shows that all keys generation strategies give a FNR at 0, except for the Bernoulli distribution with nearest interpolation at $d=16$.

\vspace{-0.5cm}

\begin{figure}[h]
    \centering
    \includegraphics[width=0.483\textwidth]{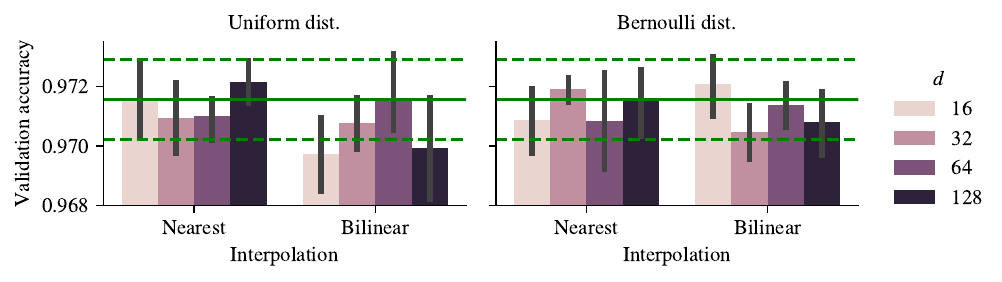}
    \caption{Validation accuracies of the poisoned AST model on Speechcommands. Green bar shows the mean accuracy of benign models, error bars show the standard deviation.}
    \label{fig:val_acc_speechcommands_ast}
\end{figure}

\begin{figure}[h]
    \centering
    \includegraphics[width=0.483\textwidth]{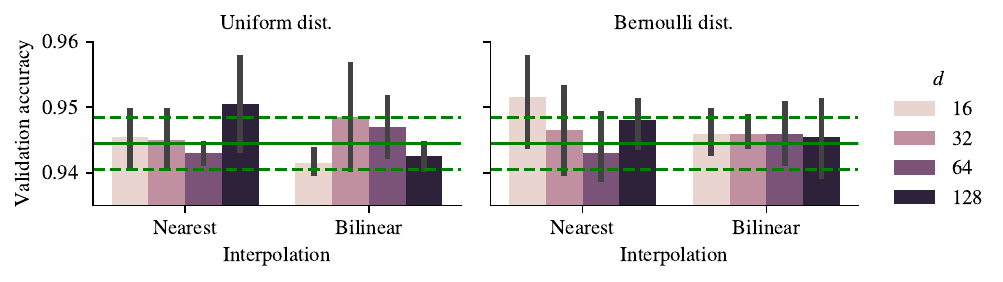}
    \caption{Validation accuracies of the poisoned AST model on ESC50. Green bar shows the mean accuracy of benign models, error bars show the standard deviation.}
    \label{fig:val_acc_esc50_ast}
\end{figure}

\paragraph*{Preserving the integrity}
To ensure that the data can still be used by legitimate users, we verify that the data poisoning does not affect the trained model's performance on the validation set.
Figures \ref{fig:val_acc_speechcommands_ast} and \ref{fig:val_acc_esc50_ast} show no measurable degradation of performance with our method, making it practical for protecting the data while preserving their integrity.

\begin{table}[h]
    \centering
    \caption{SNR of the data poisoning on the Speechcommands dataset and AST model for different strategies of keys generation ($d=128$). Standard deviation as error.}
    \begin{tabular}{|c|c|c|c|c|}
        \hline
        \textbf{Distribution}  & \multicolumn{2}{|c|}{\textbf{Uniform}} & \multicolumn{2}{|c|}{\textbf{Bernoulli}} \\
        \hline
        \textbf{Interpolation} & \textbf{nearest} & \textbf{bilinear} & \textbf{nearest} & \textbf{bilinear} \\
        \hline
        \textbf{SNR (dB)} & $55.8 \pm 7.6$ & $55.5 \pm 7.6$ & $55.9 \pm 7.4$ & $55.7 \pm 7.5$ \\
        \hline
    \end{tabular}
    \label{tab:snr}
\end{table}

\paragraph*{Stealthiness}
For our method to be practical, Bob should not be able to detect the perturbations.
Table \ref{tab:snr} shows the SNR of the data poisoning on the Speechcommands dataset and AST model for different strategies of keys generation for $d = 128$.
With SNR values above $55$ dB, the crafted perturbations have a negligible amplitude compared to the audio signal.
For a qualitative analysis, we listen to the resulting audio samples which reveal that some artifact are perceptible in the audio signal.
The low contamination rate $\epsilon$ would still make it difficult to find the poisoned samples.
Future work should introduce an audio perceptual loss in the crafting of the perturbations to enforce their imperceptibility.

\begin{figure}[h]
    \begin{subfigure}{0.483\textwidth}
        \includegraphics[width=\textwidth]{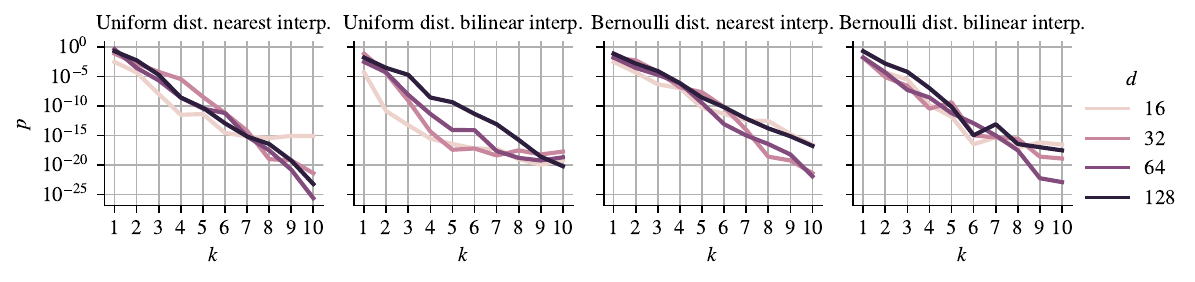}
        \caption{$p$-values.}
        \label{fig:p_values_esc50}
    \end{subfigure}
    \begin{subfigure}{0.483\textwidth}
        \includegraphics[width=\textwidth]{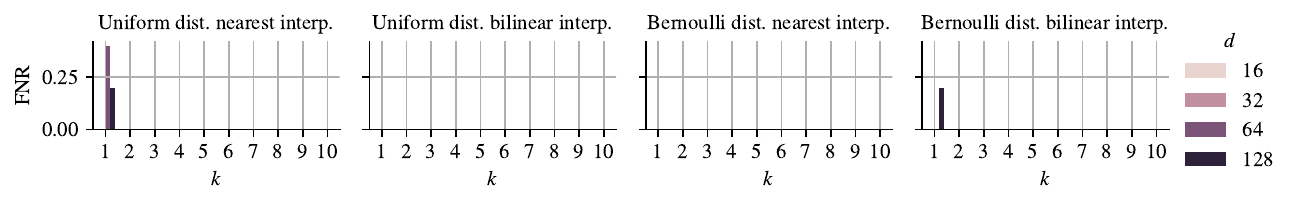}
        \caption{False Negative Rate}
        \label{fig:fnr_esc50}
    \end{subfigure}
    \caption{Results of the data poisoning on the ESC50 dataset with the AST model.}
    \label{fig:results_esc50_ast}
\end{figure}

\paragraph*{ESC50 dataset}
We also evaluate our method on the ESC50 dataset with the AST model.
Figure \ref{fig:results_esc50_ast} shows that our method similarly display high detection rate and high confidence in identifying Bob's models.
All this while preserving the integrity and utility of the data by conserving the validation accuracy as shown on Figure \ref{fig:val_acc_esc50_ast}.

\vspace{-0.5cm}

\section{Conclusion}

In this paper, we adapted a method for image datasets ownership verification to audio datasets.
With only black-box access to the model, this method allows to detect the use of the dataset with high confidence even on practical state of the art models all while preserving the model's performances.
By adapting the choice of \textit{keys} to audio datasets, we allow for a statistical test which ensures strong guarantees against false positives.
We also show the robustness of our method against common data augmentation techniques, making it a practical method to protect audio datasets.
While producing data samples with high SNR, perceivable artifacts still limit the stealthiness on a handful of samples.
Future work should leverage perceptual losses to improve the stealthiness and limit perceivable artifacts.
We believe our method can be used to protect audio datasets and ensure the integrity and ownership of the data.

\newpage
\bibliography{bibliography}
\bibliographystyle{plain}

\end{document}